\documentclass[preprint,trackchanges]{aastex62}

\usepackage{subfigure}
\usepackage{color}

\newcommand{\HI}{\ion{H}{1}}
\newcommand{\DI}{\ion{D}{1}}
\newcommand{\SiIII}{\ion{Si}{3}}
\newcommand{\kms}{km s$^{-1}$}
\newcommand{\Lya}{Ly$\alpha$}

\newcommand{\ergcmsA}{erg cm$^{-2}$ s$^{-1}$ \AA$^{-1}$}

\begin{document}

\title{FUMES. II. \Lya~Reconstructions of Young, Active M Dwarfs
        }

\author{Allison Youngblood}
\affiliation{Laboratory for Atmospheric and Space Physics, University of Colorado, 600 UCB, Boulder, CO 80309, USA}
\affiliation{NASA Goddard Space Flight Center, Greenbelt, MD 20771, USA}
\email{allison.youngblood@lasp.colorado.edu}

\author{J. Sebastian Pineda}
\affiliation{Laboratory for Atmospheric and Space Physics, University of Colorado, 600 UCB, Boulder, CO 80309, USA}

\author{Kevin France}
\affiliation{Laboratory for Atmospheric and Space Physics, University of Colorado, 600 UCB, Boulder, CO 80309, USA}
\affiliation{Department of Astrophysical and Planetary Sciences, University of Colorado, UCB 389, Boulder, CO 80309, USA}
\affiliation{Center for Astrophysics and Space Astronomy, University of Colorado, 389 UCB, Boulder, CO 80309, USA}

\begin{abstract}

The \HI\ \Lya\ (1215.67 \AA) emission line dominates the far-UV spectra of M dwarf stars, but strong absorption from neutral hydrogen in the interstellar medium makes observing \Lya\ challenging even for the closest stars. As part of the Far-Ultraviolet M-dwarf Evolution Survey (FUMES), the \textit{Hubble Space Telescope} has observed 10 early-to-mid M dwarfs with ages ranging from $\sim$24 Myr to several Gyrs to evaluate how the incident UV radiation evolves through the lifetime of exoplanetary systems. We reconstruct the intrinsic \Lya\ profiles from STIS G140L and E140M spectra and achieve reconstructed fluxes with 1-$\sigma$ uncertainties ranging from 5\% to a factor of two for the low resolution spectra (G140L) and 3-20\% for the high resolution spectra (E140M). We observe broad, 500-1000 \kms\ wings of the \Lya\ line profile, and analyze how the line width depends on stellar properties. We find that stellar effective temperature and surface gravity are the dominant factors influencing the line width with little impact from the star's magnetic activity level, and that the surface flux density of the \Lya\ wings may be used to estimate the chromospheric electron density. The \Lya\ reconstructions on the G140L spectra are the first attempted on $\lambda/\Delta\lambda\sim$1000 data. We find that the reconstruction precision is not correlated with SNR of the observation, rather, it depends on the intrinsic broadness of the stellar \Lya\ line. Young, low-gravity stars have the broadest lines and therefore provide more information at low spectral resolution to the fit to break degeneracies among model parameters.

\end{abstract}

\keywords{M dwarf stars, stellar atmospheres, interstellar absorption, planet-hosting stars}

\section{Introduction} \label{sec:Introduction}

Far-ultraviolet (FUV) photons (912-1700 \AA) drive photochemistry and heating in planetary upper atmospheres due to the large, wavelength-dependent absorption cross sections of molecules throughout the FUV (e.g., \citealt{Segura2005,Loyd2016}). Using the \textit{Hubble Space Telescope} (\textit{HST}), the Far-Ultraviolet M-dwarf Evolution Survey (FUMES; HST-GO-14640) has measured the FUV spectral energy distributions of early-to-mid M dwarfs ranging in age from 24 Myr to field age ($\sim$5 Gyr) to determine how stellar magnetic activity evolves with age and to better inform exoplanet atmosphere evolution studies (Pineda et al. \textit{accepted}). In particular, a large ratio of incident FUV to near-ultraviolet (NUV; 1700-3200 \AA) flux on a planet can lead to the abiotic production of  oxygen and ozone, possible biosignatures (see reviews by \citealt{Meadows2018} and \citealt{Schwieterman2018}). M dwarfs have intrinsically faint FUV and NUV emission from their cool photospheres, but high levels of magnetic heating (e.g., non-radiative heating) make bright chromospheric and transition region emission lines that raise the FUV/NUV flux ratio two to three orders of magnitude higher than for solar-type stars.

\HI~\Lya~(1215.67 \AA) is the brightest M dwarf emission line in the UV \citep{France2013}, and is therefore required for a thorough accounting of the stellar UV energy budget. Yet, neutral hydrogen gas in the ISM completely attenuates the inner $\sim$80--100 \kms~of the \Lya~line core for even the closest stars. To determine the intrinsic stellar emission, the \Lya~line must be reconstructed from the observed wings. Historically, this has been done at high spectral resolving power ($\lambda$/$\Delta\lambda>$40,000) so that the \DI~absorption line (-82 \kms~from \HI) can be resolved from the \HI~absorption line (e.g., \citealt{Wood2005}). Resolving the optically thin \DI~line places strong constraints on the properties of the highly optically thick \HI~line: column density, radial velocity, and Doppler broadening. \cite{France2013} showed that reliable reconstructions can be performed at lower resolving power ($\lambda$/$\Delta\lambda\sim$10,000). For the first time, we present reconstructions at an even lower spectral resolving power ($\lambda$/$\Delta\lambda\sim$1,000), where the \HI~absorption trough is completely unresolved.

The higher sensitivity of the G140L spectra compared to higher resolution STIS gratings (G140M, E140M, and E140H) eases the detection of the important \Lya~line, expanding the volume of M dwarfs for which \Lya~emission can potentially be studied. Higher sensitivity also allows for the measurement of very broad \Lya~wings ($\sim$500-1000 \kms), which have been long known for the Sun \citep{Morton1961} and for M dwarfs \citep{Gayley1994,Youngblood2016b}. The broad wings are the result of partial frequency redistribution, which occurs because \Lya~is a highly optically thick resonance line \citep{Milkey1973,Basri1979}. Photons from the lower opacity lower transition region escape in the line core, whereas photons from the higher opacity chromosphere must diffuse out into the broad wings to escape. Matching the observed wing strength of emission lines like \Lya~is a notoriously difficult problem for stellar models, especially for M dwarf models (see \citealt{Fontenla2016,Peacock2019a,Tilipman2021}). More detailed observational constraints support the upcoming generation of stellar models that include chromospheres and transition regions (\citealt{Peacock2019b,Tilipman2021}).

The intensity of the chromospheric emission line wings compared to the line core is controlled primarily by the pressure scale height \citep{Ayres1979}, with an inverse dependence on surface gravity. For main sequence stars, this means that more massive stars have brighter wings, as shown by \cite{Wilson1957} for \ion{Ca}{2} H\&K. However, there is likely a small dependence on magnetic activity \citep{Ayres1979,Gayley1994}, with more active stars exhibiting stronger wings. Combining the young, active M dwarf sample of FUMES, two well-known active M dwarfs from the literature (Proxima Centauri and AU Mic), and the more inactive M dwarf sample from the MUSCLES Treasury Survey \citep{France2016,Youngblood2016b,Loyd2016}, we address the magnitude of magnetic activity's effect on the observed \Lya~wing strength of M dwarf stars.

In Section~\ref{sec:ObservationsReductions}, we briefly describe the FUMES observations and reductions, and in Section~\ref{sec:reconstructions} we thoroughly describe the \Lya~reconstructions. These results are used in the main FUMES analysis (Pineda et al. \textit{accepted}). In Section~\ref{sec:Discussion}, we analyze the broad \Lya~wings and the implications for understanding M dwarf atmospheres. In Section~\ref{sec:Summary}, we summarize our findings.

\section{Observations and Reductions} \label{sec:ObservationsReductions}

Using the STIS spectrograph onboard \textit{HST}, we observed 10 M dwarfs as part of the The Far-Ultraviolet M-dwarf Evolution Survey (FUMES) survey (GO 14640; PI: J. S. Pineda). Properties of the targets are listed in Table~\ref{table:targs} and discussed in more detail in Paper I (Pineda et al. \textit{accepted}). Two of the targets, LP 55-41 and G 249-11, were detected at low signal-to-noise ratio (SNR), and we do not attempt \Lya~reconstructions for them. Custom reductions were performed using \texttt{stistools}\footnote{https://stistools.readthedocs.io/en/latest/} following \cite{Loyd2016}, including the exclusion of flares from the extracted spectra of GJ 4334, GJ 410, and HIP 17695. See Pineda et al. (\textit{under review}) for more details.

\begin{deluxetable}{cccccccccc}
\tablecolumns{9}
\tablewidth{0pt}
\tablecaption{ FUMES Targets  \label{table:targs} } 
\tablehead{\colhead{Name} & 
                  \colhead{Other} &
                  \colhead{Spectral} & 
                  \colhead{d} &
                  \colhead{P$_{rot}$} &
                  \colhead{M} &
                  \colhead{R} &
                  \colhead{T$_{eff}$} & 
                  \colhead{Age} &
                  \colhead{STIS} \\ 
                  \colhead{} & 
                  \colhead{Name} & 
                  \colhead{Type} & 
                  \colhead{(pc)} & 
                  \colhead{(d)} & 
                  \colhead{(M$_{\odot}$)} &
                  \colhead{(R$_{\odot}$)} & 
                  \colhead{(K)} & 
                  \colhead{} & 
                  \colhead{grating}
                  }
\startdata
G 249-11 & & M4 & 29.14 & 52.8$^a$ & 0.24 & 0.26 & 3277 & field$^e$ & G140L\\
HIP 112312 & WW PsA & M4.5 & 20.86	 & 2.4$^b$ & 0.25 & 0.69 & 3173 & 24 Myr$^f$ & E140M \\
GJ 4334 & FZ And & M5 & 25.33 & 23.5$^a$ & 0.29 & 0.31 & 3260 & field$^g$ & G140L\\
LP 55-41 & & M3 & 37.04 & 53.4$^a$ & 0.41 & 0.42 & 3412 & field$^e$ & G140L\\	
HIP 17695 & & M4 & 16.8 & 3.9$^b$ & 0.44 & 0.50 & 3393 & 150 Myr$^f$ & E140M\\
LP 247-13 & & M3.5 & 35.04 & 1.3	$^c$ & 0.50 & 0.49 & 3511 & 650 Myr$^h$ & G140L\\
GJ 49 & & M1 & 9.86 & 18.6$^d$ & 0.54 & 0.53 & 3713 & field$^i$ & G140L\\
GJ 410 & DS Leo & M0 & 11.94 & 14.0$^d$ & 0.56 & 0.55 & 3786 & 300 Myr$^h$ & G140L\\
CD -35 2722 & & M1 & 22.4 & 1.7$^b$ & 0.57 & 0.56 & 3727 & 150 Myr$^f$ & G140L\\
HIP 23309 & & M0 & 26.9 & 8.6$^b$ & 0.79 & 0.93	 & 3886 & 24 Myr$^f$ & G140L\\
\enddata
\tablecomments{Distances (d) from Gaia Data Release 2 \citep{Brown2018}; spectral types, effective temperatures (T$_{eff}$), masses (M), and radii (R) from Pineda et al. \textit{accepted}.}
\tablereferences{(a) \cite{Donati2008}, (b) \cite{Hartman2011}, (c) \cite{Messina2010}, (d) \cite{Newton2016}, (e) \cite{Gagne2018}, (f) \cite{Bell2015}, (g) \cite{Irwin2011}, (h) \cite{Shkolnik2014}, (i) \cite{Miles2017}.}
\end{deluxetable}

\section{\Lya~Reconstructions} \label{sec:reconstructions}

\subsection{The Model} \label{subsec:model}

Our model is comprised of two components: the stellar emission component and the ISM absorption component. We tested different functions for the intrinsic stellar emission, including multiple, superimposed Gaussians, and found that a single Voigt profile in emission fits both the line core and the broad wings best. We use the \texttt{astropy} \texttt{Voigt1D} function, which is based on the computation from \cite{McLean1994}. We assume no self-reversal because past results have shown that the \Lya~self-reversal of M dwarfs is small \citep{Wood2005,Guinan2016}, if present at all \citep{Youngblood2016b,Bourrier2017,Schneider2019}. Given that the \Lya~line center, the region in the spectrum where the self-reversal appears, is usually entirely hidden by the ISM and not well-constrained by the reconstruction, we assume no self-reversal is present. The Voigt emission line model component has four free parameters:

\begin{equation} \label{eq:emission}
	F_{emission}^{\lambda} = \mathcal{V}(\lambda,V_{radial}, A, FWHM_{L}, FWHM_{G}),
\end{equation}

\noindent where $V_{radial}$ is the radial velocity of the emission line (km s$^{-1}$), A is the Lorentzian amplitude (erg cm$^{-2}$ s$^{-1}$ \AA$^{-1}$; note that we parameterize it in all tables as log$_{10}$ A), and FWHM$_L$ and FWHM$_G$ (\kms) are the full-width at half maximum values for the Lorentzian and Gaussian components, respectively. For use with \texttt{Voigt1D}, $V_{radial}$, FWHM$_L$, and FWHM$_G$ are converted to \AA. For the reconstructions on the E140M spectra where \Lya~and Si III are not blended, Equation~\ref{eq:emission} is used, but for the G140L spectra where the two lines are blended, $F_{emission}^{\lambda}$ = $F_{emission, HI}^{\lambda}$ + $F_{emission, SiIII}^{\lambda}$.

We assume a single ISM absorbing cloud as such low-resolution spectra (300 \kms) are not able to distinguish between $\sim$20-40 \kms~separated clouds. \cite{Youngblood2016b} demonstrated that assuming a single-velocity ISM does not significantly impact the reconstructed \Lya~flux. For the ISM component (used for \Lya~only), we model the H I and D I absorption lines each as Voigt profiles with linked parameters using the code \texttt{lyapy}\footnote{https://github.com/allisony/lyapy} \citep{Youngblood2016b}:

\begin{equation}
	F_{absorption}^{\lambda} = \mathcal{V}(\lambda,V_{HI}, log_{10} N(HI), b_{HI}) \times \mathcal{V}(\lambda,V_{DI}, log_{10} N(DI), b_{DI}).
\end{equation} \label{eq:absorption}

\noindent V$_{HI}$ is the radial velocity (km s$^{-1}$) and is assumed to be the same for both H I (1215.67 \AA) and D I (1215.34 \AA) (V$_{HI}$ = V$_{DI}$, so V$_{HI}$ is the reported parameter). log$_{10}$ N is the logarithm of the column density (cm$^{-2}$) where N(HI) and N(DI) are linked by the parameter D/H, the deuterium to hydrogen ratio: N(DI) = N(HI)$\times$D/H. D/H is fixed to 1.5$\times$10$^{-5}$ \citep{Linsky2006}, so log$_{10}$ N(HI) is the reported parameter. The Doppler parameter $b$ controls the width of the absorption line, and we link $b_{HI}$ and $b_{DI}$ so that $b_{DI}$ = $b_{HI}$/$\sqrt{2}$. $b_{HI}$ is the reported parameter. In order to reduce the number of free parameters for the G140L reconstructions, b$_{HI}$ was fixed at 11.5 \kms\ based on the standard T=8000 K ISM \citep{Wood2004,Redfield2004}.

To model the observed (attenuated) profile, we multiply the emission and absorption models (Equations~1 \& 2) and convolve with the instrument line spread function (LSF) provided by STScI\footnote{https://www.stsci.edu/hst/instrumentation/stis/performance/spectral-resolution} for the appropriate grating and slit combinations to recover the true physical parameters and account for the non-Gaussian wings of the G140L LSF:

\begin{equation} \label{eq:e140m}
	F^{\lambda} = (\mathcal{V}_{emission}\times \mathcal{V}_{absorption}) \circledast LSF.
\end{equation}

\subsection{Fitting procedure and results} \label{subsec:fitting}

To reconstruct the \Lya~profiles, we used a likelihood-based Bayesian calculation and a Markov-Chain Monte Carlo (MCMC) method (\texttt{emcee}\footnote{https://emcee.readthedocs.io/en/latest/}; \citealt{Foreman-Mackey2013}) to simultaneously fit the model (Equation~\ref{eq:e140m}) to the observed spectra. We assume uniform (flat) priors for all parameters except for a logarithmic prior for the Doppler b value \citep{Youngblood2016b}, and a Gaussian likelihood  

\begin{equation}
	\ln \mathcal{L} = -\frac{1}{2} \sum_i^N  \frac{(y_i-y_{model,i})^2}{\sigma_{y_i}^2} + \ln(2\pi \sigma_{y_i}^2), 
\end{equation} \label{eq:likelihood}

\noindent where N is the total number of spectral data points $y_i$ with associated uncertainties $\sigma_{y_i}$, and $y_{model,i}$ corresponds to Equation~\ref{eq:e140m}. We maximize the addition of $\ln \mathcal{L}$ and the logarithm of our priors with \texttt{emcee}. We used 50 walkers, ran for 50 autocorrelation times ($\sim$10$^{5}$-10$^{6}$ steps), and removed an appropriate burn-in period based on the behavior of the walkers.

\begin{figure}
   \begin{center}
   
     \subfigure{
          \includegraphics[width=\textwidth]{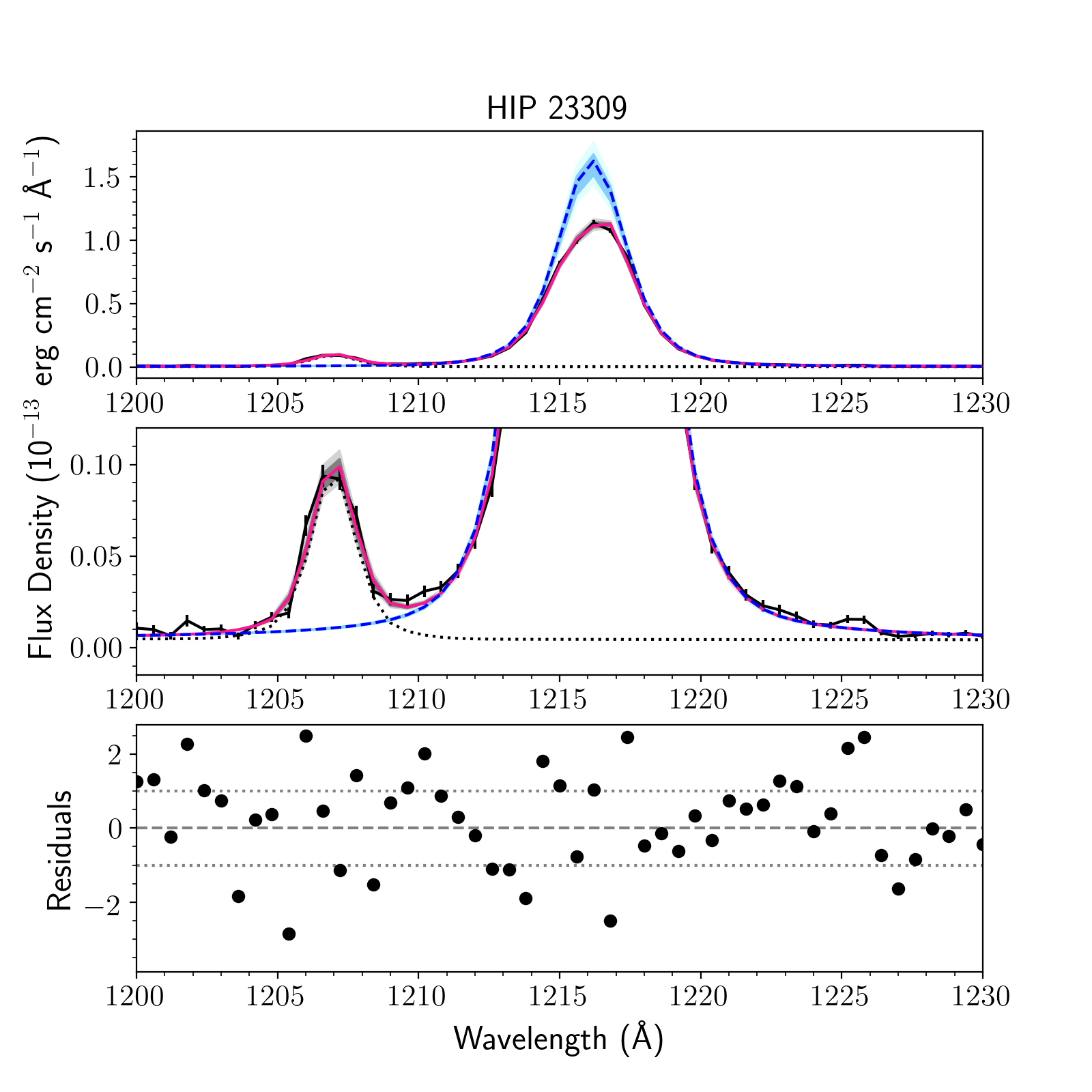}
          }

   \end{center}
    \caption{HIP 23309 best fit Ly$\alpha$~reconstruction. In the upper two panels, the STIS data with 1-$\sigma$~error bars are shown in black, the best model fit (intrinsic \Lya~profile folded through the ISM) is shown in pink with 1$\sigma$~error bars shown in dark shaded gray and 2$\sigma$~error bars in light shaded gray. The dashed blue line shows the best fit intrinsic Ly$\alpha$~profile with 1- and 2-$\sigma$~error bars (dark and light shaded blue, respectively), and the dotted black line shows the Si III best fit profile. The bottom panel shows the residuals ((data-model)/(data uncertainty)) for the best fit model (pink in the upper panels) that best fits the data (black in the upper panels). The horizontal dashed line is centered at zero, and the dotted lines are centered at $\pm$1.  
        }
    \label{fig:HIP23309_LyA_bestfit}

\end{figure} 

\begin{figure}
   \begin{center}
   
     \subfigure{
          \includegraphics[width=\textwidth]{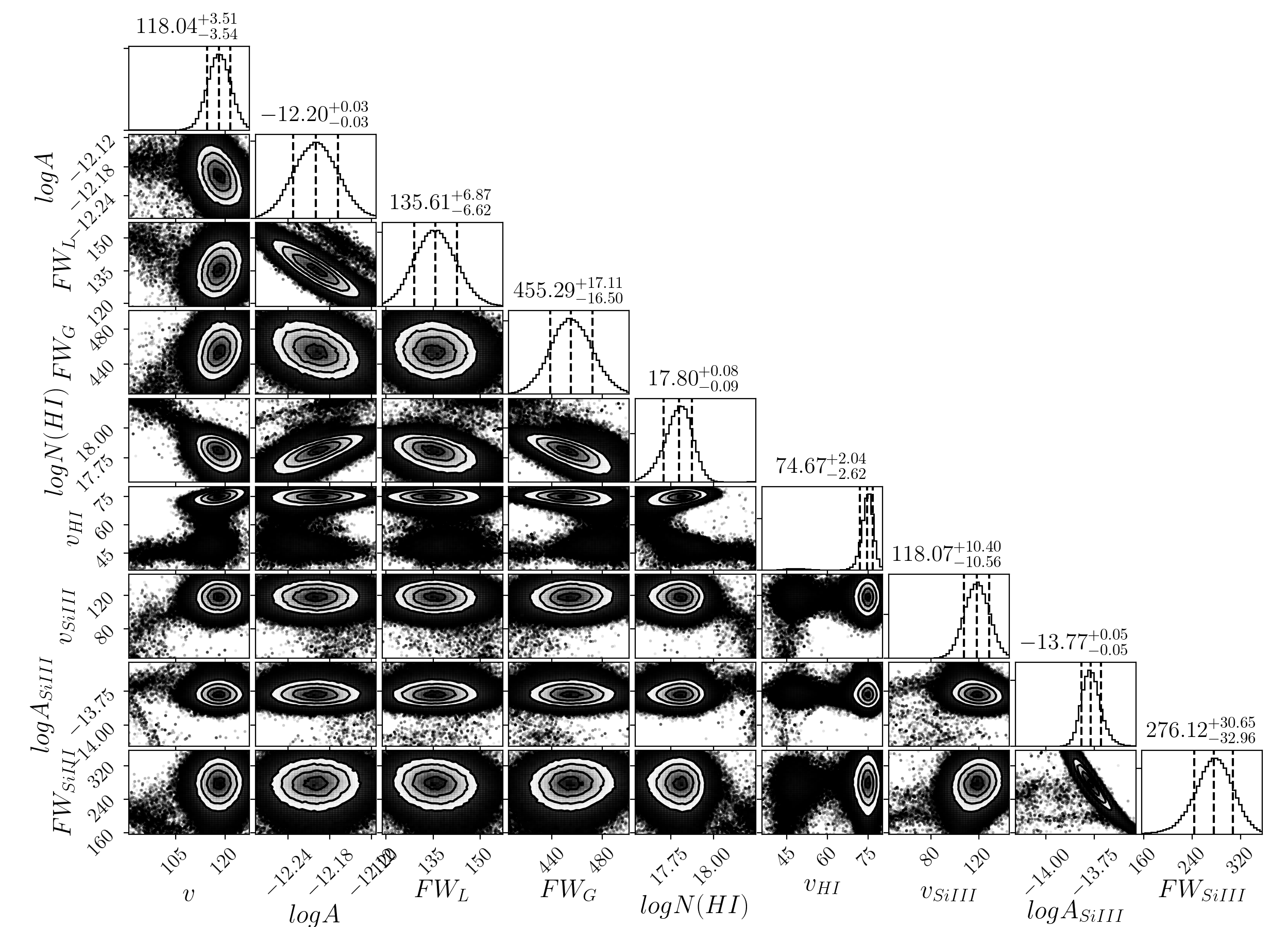}
          }

   \end{center}
    \caption{One- and two-dimensional projections of the sampled posterior probability distributions, referred to as marginalized and joint distributions, respectively, of the nine parameters for HIP 23309. Contours in the joint distributions are shown at 0.5-, 1-, 1.5-, and 2-$\sigma$, and the histograms' dashed black vertical lines show the 16th, 50th, and 84th percentiles of the samples in each marginalized distribution. The text above each histogram shows the median $\pm$~the 68\% confidence interval.
        }
    \label{fig:HIP23309_cornerplot}

\end{figure}  

Tables \ref{table:priors1}-\ref{table:priors_hip17695} show all of our model parameters with the assumed priors (uniform or logarithmic) within a bounded range and the 2.5, 15.9, 50, 84.1, and 97.5 percentiles as determined from the marginalized posterior distributions. We present the median (50th percentile) as the best fit parameter values. The best fit (median) and 68\% and 95\% confidence intervals on the reconstructed \Lya~and \ion{Si}{3} fluxes were determined from the entire ensemble (i.e., a histogram of all the \Lya~or \ion{Si}{3} fluxes from the MCMC chain). Often the median parameter values do not create a self consistent solution, so we obtain the best fit models and reconstructed profiles from the median flux in each wavelength bin from the ensemble of models and reconstructed profiles. Figure \ref{fig:HIP23309_LyA_bestfit} shows the best fit model and reconstructed profile for the HIP 23309 data, and Figure \ref{fig:HIP23309_cornerplot} shows the marginalized and joint probability distributions of the fitted parameters for HIP 23309. Similar figures for the other stars are available in the figure set in the online journal. 

In the rest of this section, we make note of any irregularities or the source of any constraints imposed on the reconstructions on a star-by-star basis. Most of the FUMES \Lya~spectra were obtained with the low-resolution G140L STIS grating ($\lambda$/$\Delta\lambda\sim$1000), where the \HI~and \DI~ISM absorption lines are unresolved. Resolving the \DI~absorption is useful for constraining the ISM model parameters (column density, Doppler b value, and radial velocity), so we provide constraints on these parameter values with outside information when necessary to aid convergence to a best-fit solution. These constraints include stellar radial velocities from SIMBAD, predicted ISM radial velocities from the Local ISM Kinematic Calculator\footnote{http://lism.wesleyan.edu/LISMdynamics.html} \citep{Redfield2008}, predicted \HI~column densities for the local interstellar cloud (LIC)\footnote{http://lism.wesleyan.edu/ColoradoLIC.html} \citep{Redfield2000}, and measured \HI~column densities from nearby sightlines collated from \cite{Wood2005}, \cite{Youngblood2016b}, and \cite{Youngblood2017}.

\paragraph{GJ 4334} The fit had to be restricted to log$_{10}$ N(HI) $>$ 17.8, because the fit preferred a log$_{10}$ N(HI) $<$17.8 solution. The likelihood values are not higher at log$_{10}$ N(HI) $<$ 17.8, but the parameter space is much more well-behaved (i.e., smoothly varying), which is likely why the fit prefers this parameter regime. With log$_{10}$ N(HI) restricted to lie between 17.8-19, the best fit log$_{10}$ N(HI) = 18.03 is in agreement with the LIC model log$_{10}$ N(HI) = 18.04 prediction and measurements of nearby sightlines (log$_{10}$ N(HI)=17.9-18.5).

\paragraph{HIP 17695} Despite the high spectral resolution obtained for this target, the fit is not consistent with probable log$_{10}$ N(HI) values ($>$17.5). The MCMC prefers the log$_{10}$ N(HI) value low ($<$17.0), which may be unphysically low based on knowledge of the local ISM \citep{Wood2005}, although a value $<$18.0 is justified based on literature measurements of nearby sightlines. We constrain the column density to be between 17.8-18.0, in agreement with the LIC model's predictions log$_{10}$ N(HI) = 17.93, and allow the MCMC to pile up near the lower boundary. We note that \ion{O}{5} (1218.3 \AA) is clearly detected in the \Lya~red wing.
 
 \paragraph{LP 247-13} We constrain the log$_{10}$ N(HI) parameter to be between 18.3-19.0 (the fit prefers $<$18.0) based on a previous measurement of log$_{10}$ N(HI) = 18.31 for a foreground star \citep{Dring1997}. 

\paragraph{GJ 49} The fit reveals 4 different local maxima with no clear global maximum. We discard the solutions with a low log$_{10}$ N(HI) = 17.7 value and a high log$_{10}$ N(HI) = 18.7 value, because nearby sightlines indicate log$_{10}$ N(HI) = 18.0-18.3. We also rule out the solution with the $>$100 \kms~difference between V$_{HI}$ and V$_{radial}$. With these restrictions on N(HI) and V$_{HI}$ in place (see priors in Table~\ref{table:priors1}), we ran the MCMC for the presented solution.

\paragraph{GJ 410} The posterior distribution for this star's fit is wide, as the 95\% confidence interval spans a factor of 15 in \Lya~flux. Nearby sightlines indicate log$_{10}$ N(HI) lies in the range of 17.6-18.6, and the solution's log$_{10}$ N(HI) = 18.32 is in agreement with this range.

\subsection{Analysis of the reconstruction quality} \label{subsec:quality_control}

The quality of the E140M reconstructions is high, but for many of our G140L reconstructions, $>$32\% of the residuals lie outside of the $\pm$1-$\sigma$ range (Figure~\ref{fig:HIP23309_LyA_bestfit} and the extended figure set in the online journal). This indicates either that the data uncertainties are underestimated or that the model is misspecified. In general, the data appear well-fit by the model, but a Durbin-Watson test \citep{Durbin1950} reveals some positive autocorrelation in the residuals. For half of our stars (HIP 23309, GJ 410, LP 247-13, HIP 112312), the Durbin-Watson statistic ($dw$) is between 1.5-1.8 (where 2 represents no autocorrelation and 0 represents perfect positive autocorrelation) and for the others (GJ 49, CD -35 2722, GJ 4334, HIP 17695) $dw$ = 1.1-1.4. This autocorrelation of the residuals can be partially accounted for by a group of weak, unresolved emission lines present around 1190-1210 \AA\ that are not included in our model. Based on detailed spectra of the Sun \citep{Curdt2001} and prominent lines in high quality M dwarf spectra like AU Mic \citep{Pagano2000,Ayres2010} and GJ 436 \citep{dosSantos2019}, these lines include \ion{S}{3} (1190, 1194, 1201, 1202 \AA), \ion{Si}{2} (1190, 1193, 1194, 1197 \AA), \ion{N}{1} (1200, 1201 \AA), \ion{Si}{3} (1206 \AA), and H$_2$ (1209 \AA)\footnote{Note that in \cite{dosSantos2019}, this line is labeled as \ion{Si}{4}, but is most likely H$_2$ as labeled in the SUMER solar spectral atlas \citep{Curdt2001}.}. There are fewer unresolved emission lines in the blue wing of the \Lya\ line, including \ion{O}{5} (1218 \AA) and \ion{S}{1} (1224, 1229, 1230 \AA). This creates an apparent asymmetry (see the solar spectrum from \citealt{Woods1995}), whereas our model is symmetric about the line center. We have included only the strongest of these adjacent emission lines (\ion{Si}{3} at 1206 \AA) in our model as the others are ill-constrained by our spectra.

We have tested adding a scatter term ($f$) to our model to account for underestimated data uncertainties, which is implemented by replacing the $\sigma_i^2$ terms in Equation 4 with $\sigma_i^2$ + $f^2$. We find that for our higher quality fits (e.g., HIP 23309), the fitted result was the same. For our lowest quality fit (GJ 49), there was a large difference in the reconstructed flux, but the quality of the fit was not improved as the structure in the residuals remained. Therefore, we do not present the fits with the scatter term in this work. We conclude that the model is missing a component, such as the weak emission lines and/or continuua in the line wings mentioned above.

GJ 49's reconstruction quality is the worst of our sample; and we note that its reconstructed \Lya\ flux should be interpreted with caution. We postulate that the reason this star's fit is so unconstrained is because of its \Lya\ line's narrow intrinsic width (see Section~\ref{subsec:G140L_future}) and the high SNR of its spectrum. GJ 49's observed spectrum has higher SNR around the \Lya\ line than any of our other G140L spectra, and this precision increases the visibility of features not covered by our model. Other FUMES targets with wider intrinsic line widths (and lower SNR) may swamp the signals from unresolved emission lines and/or continuua. Despite large scatter in the residuals, GJ 49's \Lya\ and \ion{Si}{3} flux measurements appear to be consistent with other FUMES targets of similar rotation period (Pineda et al. \textit{accepted}).

Regarding our fitted radial velocity parameters, we note that the relative accuracy of the STIS MAMA's wavelength solution is reported in the STIS Instrument Handbook as 0.25-0.5 pixels (37-74 \kms~for the G140L grating; 0.8-1.6 \kms~for the E140M grating), and the absolute wavelength accuracy is 0.5-1 pixel (74-148 \kms~for G140L; 1.6-3.3 \kms~for E140M). We find that the quoted relative wavelength accuracy can easily describe the offsets between our fitted \HI~and \SiIII~radial velocites (accounting for the 68\% confidence interval on those values). The quoted absolute wavelength accuracy can account for almost all of the offsets between the literature stellar radial velocities and our fitted radial velocities. The exception is GJ 4334, which has some disagreement in the literature over its radial velocity (-40$\pm$4 \kms~from \citealt{Newton2014}; -16.5$\pm$4.0 \kms~from \citealt{Terrien2015}; -11.9 \kms~from \citealt{West2015}). This discrepancy is not large enough to account for the $\sim$200-300 \kms~offset between our fitted radial velocities and the literature values. However, GJ 4334's velocity difference between the fitted radial velocity and the fitted ISM radial velocity is in agreement with the velocity difference of the \cite{Newton2014} radial velocity and predicted ISM velocity (6.0$\pm$1.4 \kms; \citealt{Redfield2008}), lending confidence to our fit and supporting the possibility that the absolute wavelength accuracy for GJ 4334's STIS observation is poorer than is typical. 

To test the accuracy of the reconstructions based on the G140L spectra, we degraded the resolution of our E140M spectra (HIP 112312 and HIP 17695) to the resolution of the G140L spectra by convolving with the G140L LSF and rebinning to match the G140L dispersion. Tables \ref{table:priors_hip112312}-\ref{table:priors_hip17695} show the results of the E140M (native resolution) and degraded resolution reconstructions for these two stars. There is substantial overlap between the native and degraded reconstructed \Lya~fluxes at the 68\% (for HIP 17695) and the 95\% confidence interval (for both). The uncertainties with the G140L-quality reconstruction are much larger than for the E140M reconstructions, as expected. When comparing the individual fitted parameter values, we find that the G140L-quality reconstructions do not always agree with their higher resolution counterparts. For HIP 17695, agreement between the individual fitted parameters is generally good, but not for HIP 112312. We provide confidence intervals for all of our G140L reconstruction parameters (Tables~\ref{table:priors1}-\ref{table:priors3}), but note that they should be interpreted with caution and may not reflect the true parameters that could be revealed with higher-resolution spectra.  This may be because the G140L posterior distributions are generally very wide, and we report the median parameter values as the best-fit values, even though combining the median parameter values does not always yield a self-consistent best-fit to the data. However, this exercise in comparing E140M reconstructions with degraded resolution reconstructions shows that the reconstructed \Lya~fluxes overlap within at least the 95\% level.

\begin{figure}
   \begin{center}
   
     \subfigure{
          \includegraphics[width=0.85\textwidth]{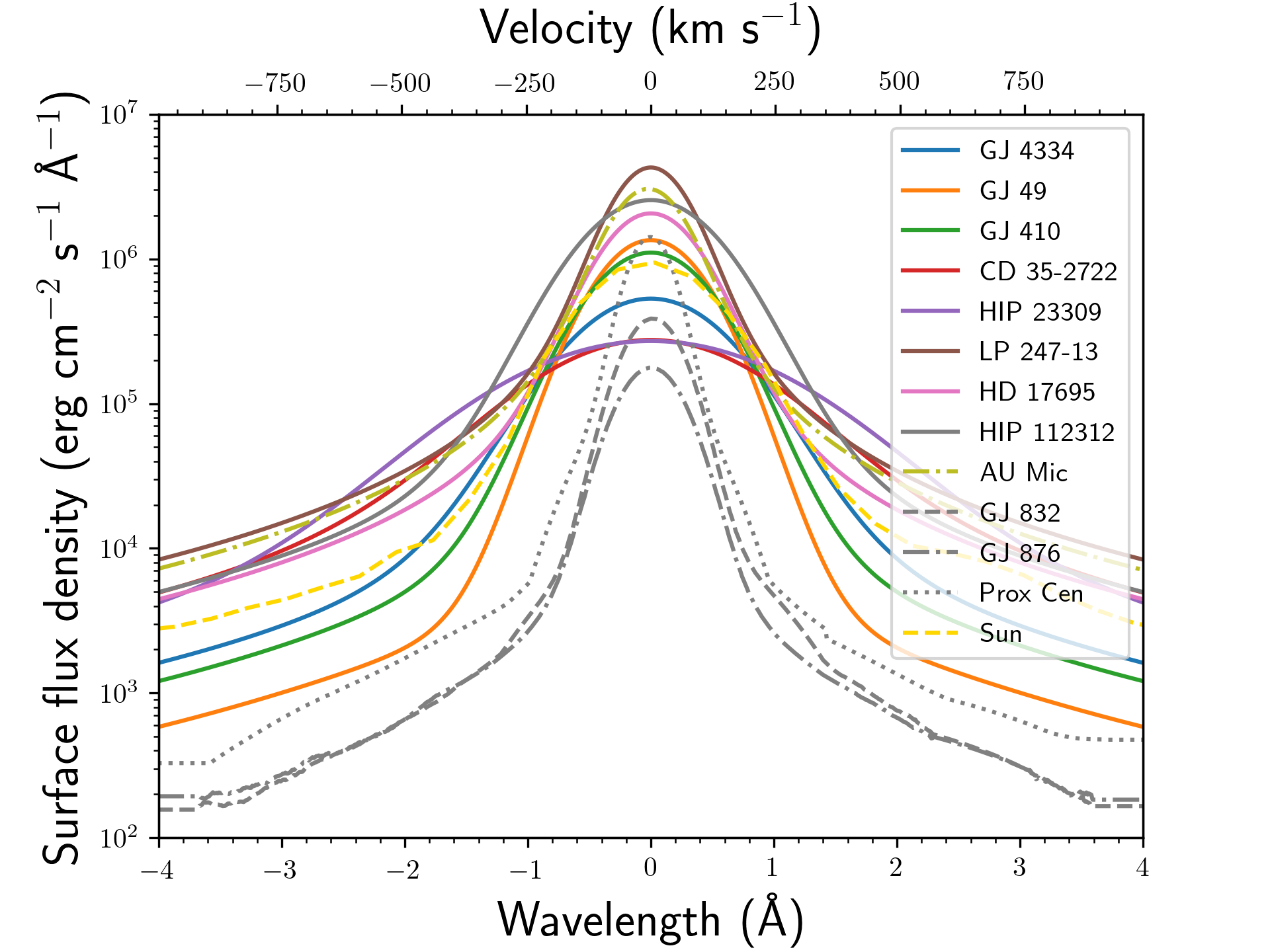}
          }

   \end{center}
    \caption{Reconstructed \Lya~profiles (corrected for stellar distance and radius) with the effect of instrumental broadening removed (with the exception of the Sun, whose line profile inside $\pm$2 \AA~is dominated by instrumental broadening) are shown for the FUMES sample and several comparison stars in dashed/dotted lines including the Sun \citep{McClintock2005}, AU Mic, Proxima Centauri, GJ 832, and GJ 876 \citep{Youngblood2016b}. The profiles have been shifted in wavelength to set the peak emission at 0 \kms.}
    \label{fig:LyA_surface_flux_spectra}

\end{figure}

\section{Discussion} \label{sec:Discussion}

\subsection{The Wilson-Bappu effect and \Lya~line widths} \label{subsec:WilsonBappu}

\begin{figure}
   \begin{center}
   
     \subfigure{
          \includegraphics[width=0.85\textwidth]{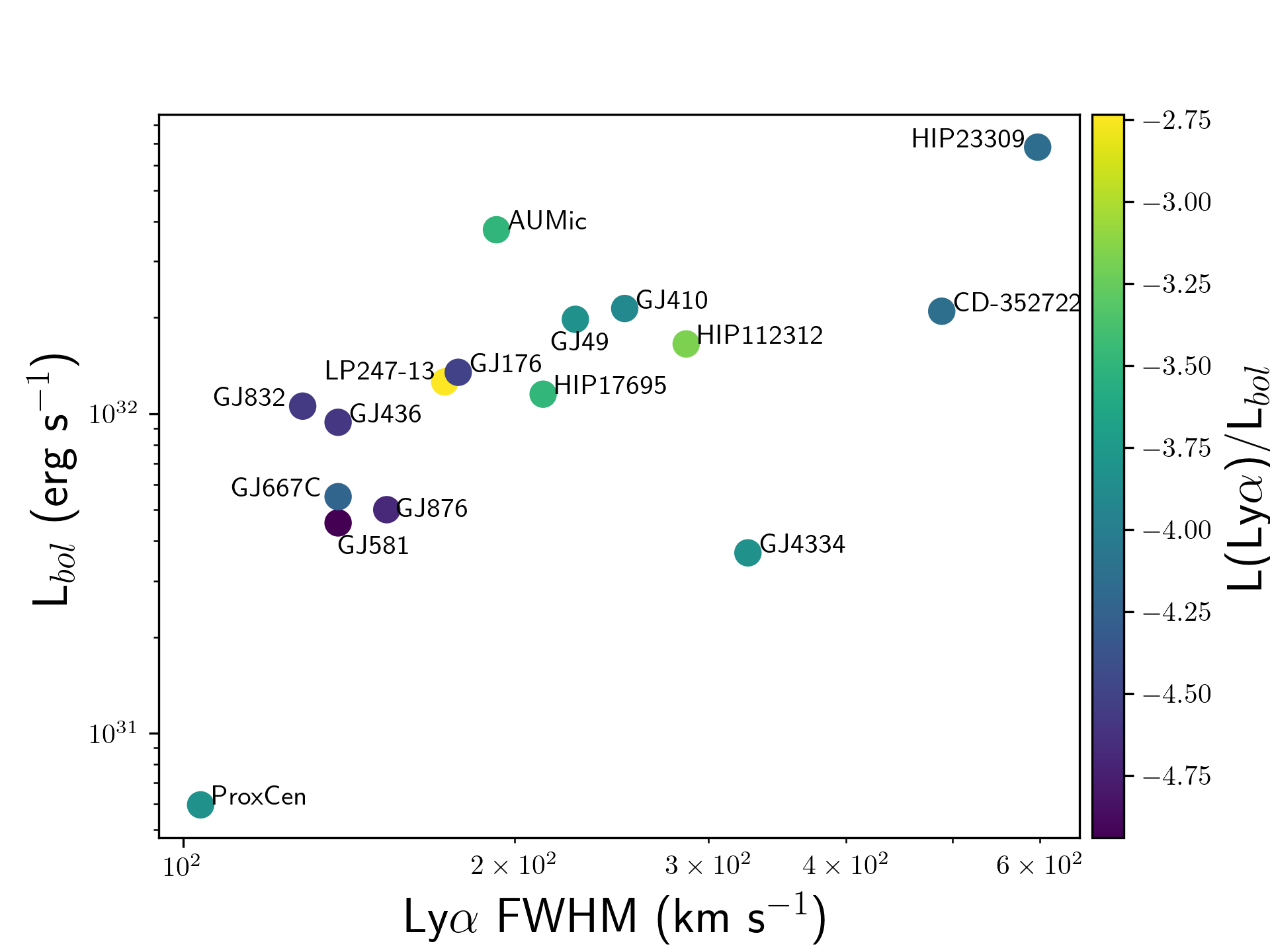}
          }

   \end{center}
    \caption{Intrinsic \Lya~FWHM (km s$^{-1}$) versus stellar bolometric luminosity, plotted analogously to \cite{Wilson1957}, and color coded by \Lya~luminosity as a fraction of bolometric luminosity. The effect of instrument line broadening has been removed. The Pearson correlation coefficient ($\rho$=0.72, p=0.0015) demonstrates a statistically significant correlation over this narrow range of parameter space.
        }
    \label{fig:Wilson-Bappu}

\end{figure}

Our STIS G140L reconstructed spectra of M dwarfs show their broad, $\sim$500-1000 \kms~\Lya~wings in detail (Figure~\ref{fig:LyA_surface_flux_spectra}). As demonstrated in \cite{Ayres1979}, the widths of chromospheric emission lines like Ca II H\&K, Mg II h\&k, and \Lya~are predominantly controlled by the stellar temperature distribution rather than chromosphere dynamics or magnetic heating. This explains the remarkable Wilson-Bappu correlation between absolute stellar magnitude and FWHM for the Ca II H\&K emission cores \citep{Wilson1957} and other chromospheric emission lines \citep{McClintock1975,Cassatella2001} across many orders of magnitude of stellar bolometric luminosity. In Figure~\ref{fig:Wilson-Bappu}, we show that our data support a similar correlation ($\rho$=0.72; p=0.0015) between bolometric luminosity and \Lya~FWHM, albeit over a much smaller parameter space than explored by \cite{Wilson1957}.

\cite{Ayres1979} notes that stellar magnetic activity (e.g., due to non-radiative heating) does play a role in the widths of chromosphere emission lines, with greater activity corresponding to wider lines, in addition to the stronger influences of stellar effective temperature, surface gravity, and elemental abundance compared to hydrogen. \cite{Ayres1979} and \cite{Linsky1980} present a linear model of chromospheric emission line width as a function of chromospheric heating (i.e., activity as measured by the flux of a chromospheric emission line), effective temperature, surface gravity, and elemental abundance. To determine which stellar properties are most responsible for our observed \Lya~widths, we construct a linear model based on our observations. We select surface gravity, Si III luminosity as a fraction of bolometric luminosity (a general ``activity" proxy), and effective temperature as predictor variables. Because we are examining a hydrogen line, we do not include a metallicity term. We scale each variable (by subtracting the mean and dividing by the standard deviation), construct a correlation matrix, and calculate the eigenvalues and eigenvectors via principal component analysis (PCA) (Table~\ref{table:PCA}). Only the first two principal components (PCs) or eigenvectors have eigenvalues $>$1 or are correlated significantly ($|\rho|$~$>$ 0.5; $p<$ 0.05) with any of the predictor variables; therefore, we only include PC$_1$ and PC$_2$ in the linear model of \Lya~width.

\begin{deluxetable}{cc|cccccc|c}
\tablecolumns{9}
\tablewidth{0pt}
\tablecaption{ Principal Component Analysis Summary}  \label{table:PCA}  
\tablehead{\colhead{} & 
                  \colhead{} &
                  \multicolumn{6}{c}{Correlation coefficients with Predictor Variables} &
                  \colhead{Regression coefficients ($\beta$)} \\ 
                  \cline{3-8}
                  \colhead{Eigenvector} &
                  \colhead{} &
                  \multicolumn{2}{c}{$g$} &
                  \multicolumn{2}{c}{L(SiIII)/Lbol} &
                  \multicolumn{2}{c}{T$_{eff}$} &
                  \colhead{Response Variable} \\ 
                  \cline{3-4} \cline{5-6} \cline{7-8}
                  \colhead{(PC)} &
                  \colhead{Eigenvalue} &
                  \colhead{$\rho$} &
                  \colhead{$p$} &
                  \colhead{$\rho$} &
                  \colhead{$p$} &
                  \colhead{$\rho$} &
                  \colhead{$p$} &
                  \colhead{FW$_{20\%}$}
                  }
\startdata
PC$_1$ & 1.55 & -0.89 & 4.5$\times$10$^{-6}$ & 0.75 & 7.6$\times$10$^{-4}$ & 0.44 & 0.09 & $\beta_1$=0.60 \\
PC$_2$ & 1.02 & -0.02 & 0.95 & -0.53 & 0.04 & 0.86 & 1.8$\times$10$^{-5}$ & $\beta_2$=0.10 \\
PC$_3$ & 0.43 & -0.5 & 0.07 & -0.4 & 0.13 & -0.3 & 0.35 & -- \\
\enddata
\tablecomments{The principal components (PCs) are related to the scaled predictor variables as follows: PC$_1$ = $-$0.71 log$_{10}$ $g$ + 0.60 log$_{10}$ L(SiIII)/Lbol + 0.36 log$_{10}$ T$_{eff}$; PC$_2$ = $-$0.02 log$_{10}$ $g$ $-$ 0.52log$_{10}$ L(SiIII)/L(bol) + 0.85 log$_{10}$ T$_{eff}$; PC$_3$ = $-$0.70 log$_{10}$ $g$ $-$ 0.60 log$_{10}$ L(SiIII)/Lbol $-$ 0.38 log$_{10}$ T$_{eff}$. $\rho$ is the correlation coefficient and $p$ is the probability of no correlation between the PCs and predictor variables. We define a significant correlation as $|\rho|$~$>$ 0.5; $p<$ 0.05. The regression coefficients relate the PCs and response variable as follows: log$_{10}$ FW$_{20\%}$ = PC$_1 \times \beta_1$ + PC$_2 \times \beta_2$. The intercept coefficient on the regression is vanishingly small ($<$10$^{-15}$) and is dropped. The linear model's predicted FW$_{20\%}$ is significantly and positively correlated with the measured values ($\rho$=0.76, $p$=6.5$\times$10$^{-4}$).}
\end{deluxetable}

We perform a multiple linear regression to relate our previously determined PCs to a response variable, the \Lya\ full width at 20\%~maximum flux (FW$_{20\%}$), a term that is analogous to W(K$_1$) from \cite{Ayres1979}. Regression coefficients are reported in Table~\ref{table:PCA}. Simplifying the linear model expressions into the original unscaled predictor variables rather than PCs, we find that for the \Lya~emission line: 

\begin{equation} \label{eq:FW20}
	\log_{10} FW_{20\%} = -0.29 \log_{10} g + 0.09 \log_{10} \frac{L(SiIII)}{L(bol)} + 2.13 \log_{10} T_{eff} - 5.54,
\end{equation}

\noindent where FW$_{20\%}$ is in \AA, $g$ is in cm s$^{-2}$, L(SiIII)/Lbol is unitless, and T$_{eff}$ is in K. There are some similarities in the coefficients between this paper's Equation \label{eq:FW20}5 and Equation 8 from \cite{Linsky1980} (log W(K$_1$) = -0.25 log $g$ + 0.25 log $F$ + 1.75 log T$_{eff}$ + 0.25 log A$_{met}$, where $F$ is the scaled non-radiative heating rate and A$_{met}$ is the metal abundance), such as the sign and magnitude of each coefficient being roughly the same. Dissimilarities are likely due to the differences in terms ($F$ and $A_{met}$) and parameter ranges in the sample stars. In this analysis, the stars used have log$_{10}$ g between 3.9-5.2, log$_{10}$ L(SiIII)/Lbol between -7.5 and -5.0, and $T_{eff}$ between 3000-3900 K. The observed range of FW$_{20\%}$ values is 0.6-3.8 \AA.

As is the case for \ion{Ca}{2}, stellar activity appears to be a minor factor in the width of \Lya, also indicated by the lack of correlation between FW$_{20\%}$ and L(\Lya)/L$_{bol}$ ($\rho$=-0.06, $p$=0.82) or L(SiIII)/L$_{bol}$ ($\rho$=0.29, $p$=0.27). The more dominant factors are surface gravity and effective temperature, indicated by the correlation coefficients between FW$_{20\%}$ and T$_{eff}$ ($\rho$=0.50; $p$=0.05) or $g$ ($\rho$=-0.51; $p$=0.04). From Figure~\ref{fig:LyA_surface_flux_spectra}, we find that in general, the M dwarfs with larger \Lya~wing flux values tend to be more active. The ``inactive" MUSCLES M dwarfs (as determined by optical activity indicators such as Ca II; \citealt{France2016}) have the narrowest profiles, and Proxima Centauri has a surprisingly narrow profile given its known levels of moderate activity \citep{Robertson2013,Robertson2016,Davenport2016,Howard2018}. For example, Proxima Centauri's log$_{10}$ L(SiIII)/L$_{bol}$ = -6.2 compared to the -7.2 to -7.5 values for the inactive MUSCLES M dwarfs GJ 832, GJ 581, and GJ 436. However, as discussed, these line widths are dominated by stellar structure, and in general, lower surface gravity (i.e., young) M dwarfs tend to be more active.

\subsection{Chromospheric electron density estimates from \Lya~observations} \label{subsec:electrondensity}

The electron density in the line forming region (the chromosphere for the \Lya\ broad wings) is a main factor in controlling the width of the \Lya~line \citep{Gayley1994}. We estimate chromosphere electron density values, which are a valuable constraint for stellar models, using the formalism from \cite{Gayley1994} that explicitly relates the surface flux density of the \Lya~broad wings to chromospheric electron density and other stellar properties:

\begin{equation} \label{eq:gayley}
F_{wing}(\Delta \lambda) \approx \frac{F_{peak, \odot}}{\Delta \lambda ^{2}}~\Big(\frac{n_e}{n_{e, \odot}}\Big)^2~\frac{g_{\odot}}{g}~\frac{T_{chromo}}{7500 K}~\frac{J_{2c,\odot}}{J_{2c}},	
\end{equation}

\noindent where $F_{wing}$($\Delta \lambda$) is the \Lya~surface flux density at $\Delta \lambda$ \AA~from line center, $F_{peak, \odot}$ is the peak solar \Lya~flux ($\sim$3$\times$10$^{5}$ erg cm$^{-2}$ s$^{-1}$), $n_e$ is the chromospheric electron density, $g$ is the surface gravity, $T_{chromo}$ is the chromospheric temperature, and $J_{2c}$ is the Balmer continuum flux. Each parameter is normalized to the solar ($\odot$) value. Stars with larger electron densities and hotter chromospheres will have broader wings, but the wing intensity is diminished for stars with greater surface gravity and greater Balmer continuum flux.

\begin{figure}
   \begin{center}
   
     \subfigure{
          \includegraphics[width=\textwidth]{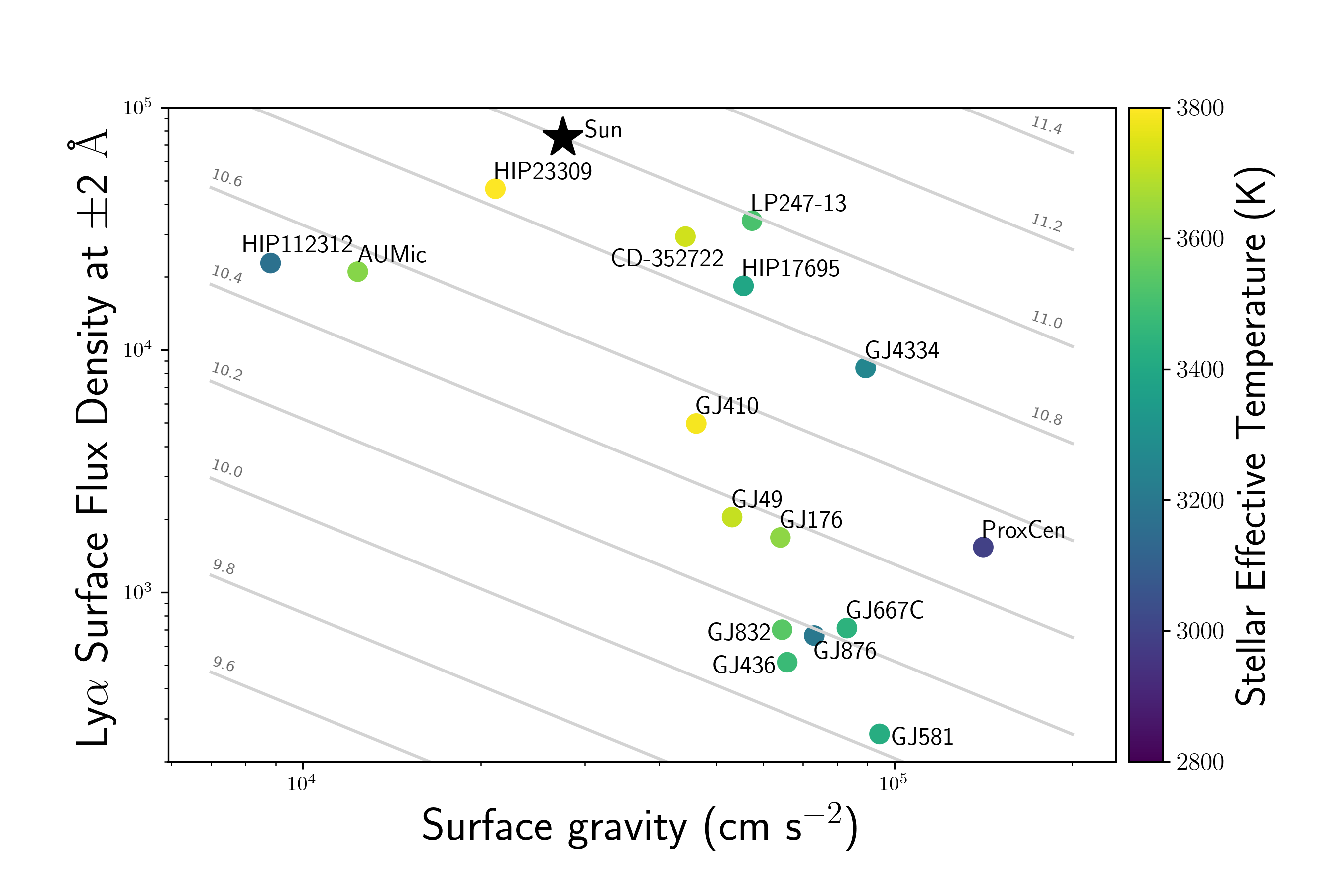}
          }

   \end{center}
    \caption{Surface gravity vs. surface flux density (erg cm$^{-2}$ s$^{-1}$ \AA$^{-1}$) at $\pm$2 \AA\ from \Lya\ line center. Lines of constant electron density (log$_{10}$ n$_e$) are shown in gray under the approximation from \cite{Gayley1994} (Equation~\ref{eq:gayley}, with T$_{chromo}$ = 7500 K, and J$_{2c}$ = J$_{2c,\odot}$). Note that the Sun's measured electron density is 10$^{11}$ cm$^{-3}$ \citep{Song2017} and GJ 832's is 10$^{10}$ cm$^{-3}$ \citep{Fontenla2016}.
        }
    \label{fig:Surface_gravity_surface_flux}

\end{figure}

Figure~\ref{fig:Surface_gravity_surface_flux} shows the \Lya~surface flux densities of the FUMES targets, the MUSCLES M dwarfs \citep{France2016}, Proxima Centauri and AU Mic \citep{Youngblood2017}, and the Sun (SORCE/SOLSTICE; \citealt{McClintock2005}), plotted against surface gravity. Lines of constant electron density are drawn on the plot using Equation~\ref{eq:gayley}. We assume $T_{chromo}$ and $J_{2c}$ are both equivalent to solar values ($T_{chromo}$ = 7500K; $J_{2c}$ = 1.7$\times$10$^{5}$ erg cm$^{-2}$ s$^{-1}$ \AA$^{-1}$ sr$^{-1}$). For stars with known chromospheric electron densities, the \citet{Gayley1994} approximation works well. The Sun's electron density log$_{10}$ $n_e$=11 cm$^{-3}$ \citep{Song2017}, and GJ 832's log$_{10}$ $n_e$=10 cm$^{-3}$ \citep{Fontenla2016}, are both in agreement with the gray curves in Figure~\ref{fig:Surface_gravity_surface_flux}. 

 We find that LP 247-13, a 625 Myr M2.7V star, has a chromospheric electron density similar to the Sun. All of the FUMES targets (``active" stars) have electron densities larger than that of the ``inactive" M dwarfs from the MUSCLES survey, except for GJ 176. We note that GJ 176 is the least ``inactive" of the MUSCLES stars as it is the most rapidly rotating (P$_{rot}$=39.5 day, \citealt{Robertson2015}) and is possibly younger than 1 Gyr based on its large X-ray luminosity \citep{Guinan2016,Loyd2018}.

\subsection{STIS G140L and future \Lya~observations} \label{subsec:G140L_future}

\begin{figure}
   \begin{center}
   
     \subfigure{
          \includegraphics[width=\textwidth]{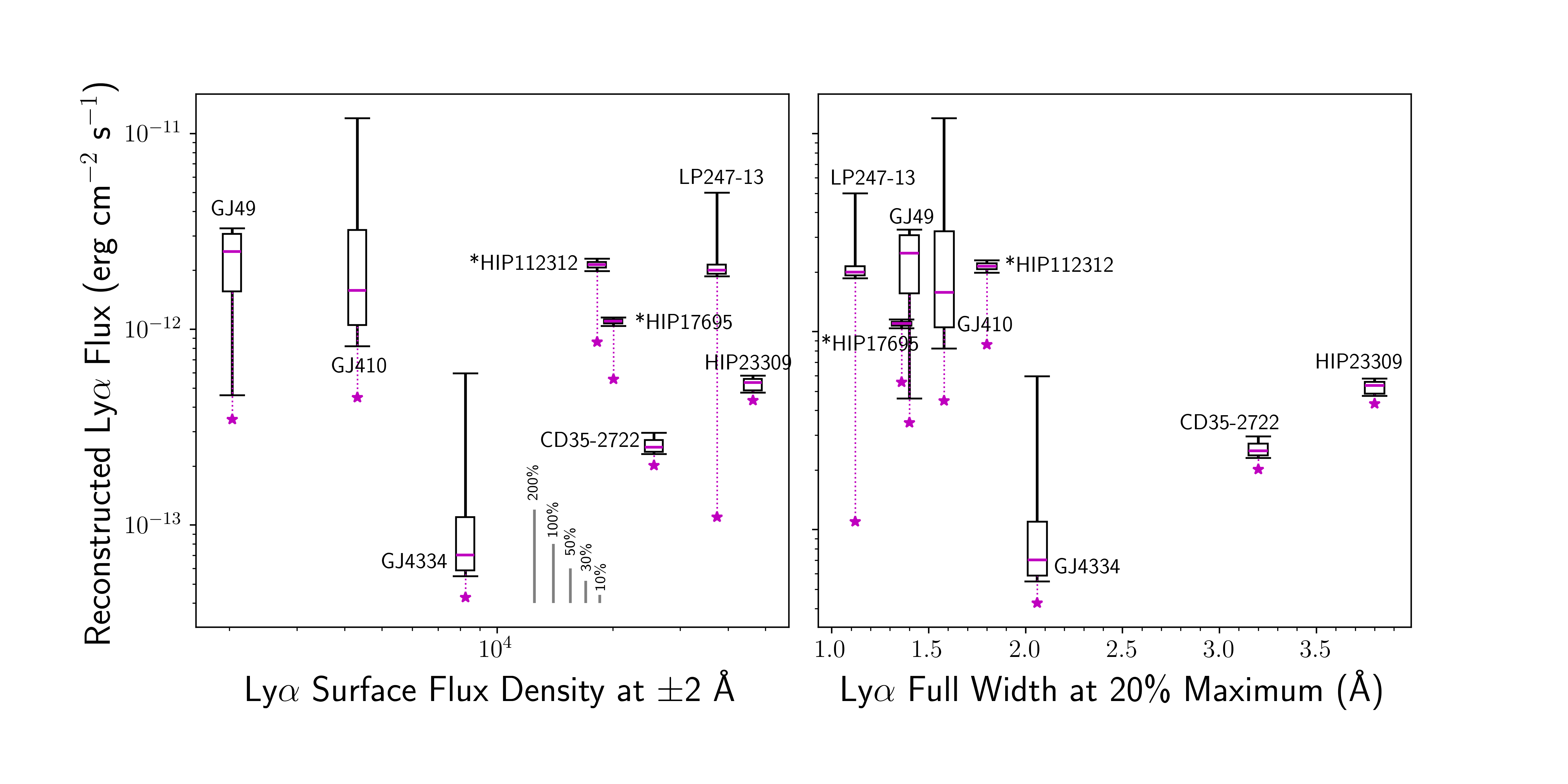}
          }

   \end{center}
    \caption{The reconstructed \Lya~flux as observed at Earth are plotted against the \Lya~surface flux density (erg cm$^{-2}$ s$^{-1}$ \AA$^{-1}$ at $\pm$2 \AA~from line center (\textit{left panel}) and \Lya~full width at 20\% maximum ($FW_{20\%}$; \textit{right panel}) for the FUMES targets. All but two stars were observed with the STIS G140L grating ($\lambda/\Delta\lambda\sim$1,000); HIP 112312 and HIP 17695 were observed with the STIS E140M grating ($\lambda/\Delta\lambda\sim$45,000) and are labeled with asterisks. The magenta horizontal line in each white box represents the median flux, the white boxes represent the 68\% confidence interval, and the black error bars represent the 95\% confidence interval. The small magenta stars show the \Lya~flux density obtained by integrating over the observed (ISM-attenuated) \Lya~spectra. The gray vertical lines indicate the apparent size of several factors of uncertainty (200\%, 100\%, 50\%, 30\%, and 10\%).}
        
    \label{fig:boxplot}

\end{figure}

The presented \Lya~reconstructions are the first based on $\lambda/\Delta\lambda\sim$1,000 spectra with the ISM \HI~absorption completely unresolved. Using the STIS G140L mode provides some observational advantages including avoiding prohibitively long exposure times of higher resolution STIS modes for M dwarf targets deemed too hazardous for the COS instrument (Bright Object Protections
\footnote{http://www.stsci.edu/hst/cos/documents/isrs/ISR2017\_01.pdf}).
 Based on the six M dwarfs presented here, we find that the precision of \Lya~reconstructions performed on STIS G140L spectra can range from 5\% to 100\% at the 68\% confidence level (Figure~\ref{fig:boxplot}). At the 95\% confidence level, the precisions range from $\sim$10\% to a factor of nine. There appears to be no dependence of these precisions on the SNR of the observed spectrum;  we note that all G140L \Lya~emission lines were detected at high SNR (90-250 integrated over the line). Rather, our three G140L targets with the largest reconstructed flux uncertainties (GJ 4334, GJ 49, and GJ 410) are also the G140L targets with the lowest surface flux in the \Lya~wings, or in other words, the narrowest profiles. We hypothesize that for narrow profiles (\Lya~surface flux at $\pm$2 \AA~$\lesssim$10$^4$ erg cm$^{-2}$ s$^{-1}$ \AA\ or $FW_{20\%}$ $<$ 2.5-3.0 \AA), the spectrum does not provide enough spectrally-resolved information for the fit to distinguish between large flux, large ISM column solutions and small flux, small ISM column solutions. The higher resolution E140M grating results in reconstructed flux precisions of approximately 2-4\% at the 68\% confidence level for high SNRs (we note that the two stars with E140M observations, HIP 112312 and HIP 17695, have a line-integrated SNR = 70-90). However, for lower SNR spectra, \cite{Youngblood2016b} found uncertainties up to 150\% in E140M reconstructions of K dwarfs (HD 97658, HD 40307, HD 85512) with SNR = 20-30 integrated over the line. The precision found by \cite{Youngblood2016b} with the STIS G140M grating ($\lambda/\Delta\lambda\sim$10,000) is 5-30\% for medium-to-high SNRs and can be a factor of $\sim$2 for low SNRs (e.g., GJ 1214, SNR=4 integrated over the line). Thus, STIS G140L spectra can produce reconstructed \Lya~fluxes for young, active M dwarfs with precisions comparable to G140M spectra, but the precision is much lower than what is obtainable with high SNR G140M or E140M spectra.

 Adopting FW$_{20\%}$ $>$ 2.5 \AA\ as the threshold between precise and imprecise \Lya\ reconstructions with G140L, Equation~5 may be useful for guiding future observers toward whether or not G140L is suitable for a \Lya\ reconstruction for a particular M dwarf. Surface gravity and effective temperature, two of the three stellar parameters in Equation~5, are readily available in the literature for many M dwarfs. The third parameter, $L$(SiIII)/$L$(bol), is not available for most M dwarfs, but can be estimated from the stellar rotation period (Pineda et al. \textit{accepted}) or common activity indicators like $R^{\prime}_{HK}$ or $L$(H$\alpha$)/$L$(bol) \citep{Melbourne2020}. 
 
 Figure~\ref{fig:boxplot} shows how the observed \Lya~fluxes compare to the reconstructed (intrinsic) fluxes. The observed fluxes were obtained simply by integrating over the observed, ISM-attenuated \Lya~profiles. In some cases, the observed \Lya~fluxes are only 10-50\% less than the reconstructed fluxes, while in others they are a factor of a few to an order of magnitude less. The dominant factor in the flux differences is the column density of the ISM absorbers and the radial velocity of the ISM absorbers relative to the stellar radial velocity. A small radial velocity offset between the star and ISM, and larger column densities will result in larger flux differences between observed and reconstructed. Figure~\ref{fig:boxplot} may give the reader a sense of whether or not  performing a reconstruction on G140L \Lya~spectra is worthwhile for their science goals.

\section{Summary} \label{sec:Summary}

As part of the Far Ultraviolet M-dwarf Evolution Survey (FUMES), we have reconstructed the intrinsic \Lya~profiles of 8 early-to-mid M dwarfs spanning a range of young to field star ages from low and moderate resolution spectra taken with \textit{HST}'s STIS spectrograph. The \Lya~and \SiIII~fluxes derived in this paper are incorporated into Paper I of the FUMES survey (Pineda et al. \textit{accepted}), which describes the flux evolution of FUV spectral lines with stellar age and rotation period for early-to-mid M dwarfs. We summarize our findings here:

\begin{enumerate}
	\item We present the first demonstration of \Lya~reconstruction on low, $\lambda/\Delta\lambda\sim$1,000 resolution spectra, where the \HI~absorption trough from the ISM is completely unresolved. We find that the 1-$\sigma$ precision in the reconstructed \Lya~flux can be 5-10\% in the best case (young M dwarfs) and a factor of two in the worst case (field age M dwarfs). The precision is not correlated with SNR of the observation, rather, it depends on the intrinsic broadness of the stellar \Lya~line. Young, low-gravity stars have the broadest lines and therefore provide more information at low spectral resolution to the fit to break degeneracies among model parameters. 
	\item Our high SNR, low resolution \Lya~spectra detect the extremely broad wings ($\sim$500-1000 \kms) at SNR=7-14 per resolution element, and we see large differences in the width of \Lya~from star to star. We confirm past findings that the line width is predominantly correlated with the fundamental stellar parameters surface gravity and effective temperature, rather than magnetic activity.
	\item \Lya~surface flux density $\sim$2 \AA~from line center may predict electron density values in the chromosphere, as shown by \cite{Gayley1994}. We confirm the validity of the \Lya~surface flux density approximation from that work using GJ 832's spectrum from \cite{Youngblood2016b,Loyd2016} and modeled electron density from \cite{Fontenla2016}.
\end{enumerate}

\acknowledgments
The data presented here were obtained as part of the \textit{HST} Guest Observing program \#14640. A.Y. acknowledges support by an appointment to the NASA Postdoctoral Program at Goddard Space Flight Center, administered by USRA through a contract with NASA. We thank J. Linsky, T. Barclay, and A. Wolfgang for helpful discussions, and E. R. Newton and W. C. Waalkes for their contributions to \texttt{lyapy}.

\facilities{HST}
\software{Astropy \citep{Robitaille2013}, IPython \citep{Perez2007}, Matplotlib \citep{Hunter2007}, NumPy and SciPy \citep{VanderWalt2011}, lyapy \citep{Youngblood2016b}, emcee \citep{Foreman-Mackey2013}, triangle \citep{Foreman-Mackey2014}, statsmodels \citep{Seabold2010}.}

\begin{deluxetable}{ccc}
\tablecolumns{3}
\tablewidth{0pt}
\tablecaption{ Prior Probabilities, Best Fits, and Confidence Intervals for G140L \label{table:priors1}} 
\tablehead{\colhead{Parameter} & 
                  \colhead{GJ 4334} &
                  \colhead{GJ 49}    
                  }
\startdata
V$_{radial}$  & U(-100; 300) &  U(-150; 150)    \\
(\kms) & [142.4, 153.0, 166.7, 178.0, 188.9] &  [43.8, 50.9, 52.8, 54.0, 55.1]  \\
\hline 
log$_{10}$ A  & U(-18.5, 8) & U(-14; -10)  \\
(\ergcmsA) & [-12.99, -12.89, -12.72, -12.36, -10.95] & [-11.93, -11.07, -10.52, -10.32, -10.26] \\
\hline
FWHM$_{L}$  & U(1; 1000) & U(1; 1000) \\
(\kms) & [8.6, 37.9, 57.5, 73.6, 88.1] &  [9.3, 10.1, 12.6, 23.7, 62.3]   \\
\hline
FWHM$_{G}$  & U(1; 1000) & U(1; 1000)  \\
(\kms) & [167.4, 209.5, 253.7, 293.4, 327.1] &  [178.8, 181.8, 187.7, 208.4, 277.2]  \\
\hline
log$_{10}$ N(\HI)  & U(17.8; 19) & U(17.7; 18.5)  \\
(cm$^{-2}$) & [17.81, 17.87, 18.03, 18.25, 18.58] & [17.81, 18.29, 18.44, 18.49, 18.50]  \\
\hline
b$_{HI}$  & 11.5 & 11.5  \\
(\kms) &  &  \\
\hline
V$_{HI}$  & U(0; 300) & U(0; 150)  \\
(\kms) & [188.9, 214.7, 223.1, 227.8, 233.4] & [47.0, 72.4, 72.8, 73.1, 73.3]  \\
\hline
V$_{\rm Si III}$ & U(-60; 400) & U(-250; 250)  \\
 & [195.5, 221.6, 247.7, 270.6, 292.7] & [76.2, 90.0, 104.2, 118.7, 132.9]  \\
 \hline
A$_{\rm Si III}$ & U(-16; -12) & U(-16; -13) \\
 & [-14.94, -14.85, -14.75, -14.61, -14.37] & [-14.55, -14.49, -14.44, -14.37, -14.30]  \\
 \hline
FWHM$_{\rm Si III}$ & U(1; 700) & U(1; 700) \\
 & [130.1, 202.1, 277.0, 344.4, 416.7] & [253.0, 300.1, 349.5, 400.6, 453.1] \\
\hline
\hline
F(\Lya)  & [5.47, 5.87, 7.03, 10.96, 59.54] & [0.46, 1.56, 2.49, 3.07, 3.28]  \\
(erg cm$^{-2}$ s$^{-1}$) & $\times$10$^{-14}$ & $\times$10$^{-12}$ \\
F(Si III)  & [1.78, 1.93, 2.11, 2.28, 2.44] &  [5.06, 5.27, 5.50, 5.71, 5.90]  \\
(erg cm$^{-2}$ s$^{-1}$) & $\times$10$^{-15}$ & $\times$10$^{-15}$ \\
\enddata
\tablecomments{U represents a uniform prior within the bounds. Other values are fixed values. On the second line: [2.5\%, 15.9\%, 50\%, 84.1\%, 97.5\%].}
\end{deluxetable}

\begin{deluxetable}{ccc}
\tablecolumns{3}
\tablewidth{0pt}
\tablecaption{ Prior Probabilities, Best Fits, and Confidence Intervals for G140L \label{table:priors2}} 
\tablehead{\colhead{Parameter} & 
                  \colhead{GJ 410} &
                  \colhead{LP247-13}   
                  }
\startdata
V$_{radial}$  & U(-250; 250) & U(-250; 250)\\
(\kms) & [0.8, 8.7, 16.8, 25.5, 36.8] &  [84.8, 91.6, 101.3, 110.2, 115.3] \\
\hline 
log$_{10}$ A  & U(-18.5; -8) &  U(-18; -8) \\
(\ergcmsA) &  [-11.6, -11.38, -11.04, -10.44, -9.3] &  [-12.27, -12.19, -12.0, -11.58, -11.04]\\
\hline
FWHM$_{L}$  & U(1; 1000) & U(1; 1000) \\
(\kms) & [3.8, 13.8, 27.3, 40.2, 51.6] & [22.6, 41.6, 66.7, 82.7, 91.4]  \\
\hline
FWHM$_{G}$  & U(1; 1000) & U(1; 1000) \\
(\kms) &  [158.4, 181.8, 204.7, 226.6, 246.6] & [17.5,  72.2, 118.1, 153.6, 178.3] \\
\hline
log$_{10}$ N(\HI)  & U(17.5; 19) & U(18.3; 19) \\
(cm$^{-2}$) & [18.05, 18.18, 18.32, 18.48, 18.68] & [18.3, 18.31, 18.35, 18.42, 18.50] \\
\hline
b$_{HI}$  & 11.5 & 11.5  \\
(\kms) &  &   \\
\hline
V$_{HI}$  & U(-200; 200) & U(-250; 250)\\
(\kms) & [71.4, 72.7, 73.6, 74.7, 76.0] & [74.4, 76.0, 78.8, 84.4, 88.4] \\
\hline
V$_{\rm Si III}$ & U(-160; 350) & U(-250; 250) \\
 (\kms) &  [-33.7, -11.8, 9.8, 31.6, 53.6] & [121.1, 143.8, 166.6, 188.7, 209.8]\\
 \hline
A$_{\rm Si III}$ & U(-16; -12) & U(-16; -13)\\
 (\ergcmsA) & [-14.28, -14.19, -14.1, -13.99, -13.75] & [-14.50, -14.43, -14.36, -14.28, -14.20] \\
 \hline
FWHM$_{\rm Si III}$ & U(1; 700) & U(1; 700) \\
 (\kms) & [113.1, 197.5, 257.5, 313.7, 374.1] & [268.3, 325.2, 391.3, 464.7, 539.7] \\
\hline
\hline
F(\Lya)  &  [0.82, 1.05, 1.58, 3.21, 11.95] & [3.06, 3.39, 4.42, 8.16, 17.17] \\
(erg cm$^{-2}$ s$^{-1}$) & $\times$10$^{-12}$ & $\times$10$^{-12}$ \\
F(Si III)  & [7.32, 8.02, 8.76, 9.48, 10.16] & [6.53, 6.91, 7.31, 7.72, 8.10] \\
(erg cm$^{-2}$ s$^{-1}$) & $\times$10$^{-15}$ & $\times$10$^{-15}$  \\
\enddata
\tablecomments{U represents a uniform prior within the bounds. Other values are fixed values. On the second line: [2.5\%, 15.9\%, 50\%, 84.1\%, 97.5\%].}
\end{deluxetable}

\begin{deluxetable}{ccc}
\tablecolumns{3}
\tablewidth{0pt}
\tablecaption{Prior Probabilities, Best Fits, and Confidence Intervals for G140L (continuation of Table~\ref{table:priors1}) \label{table:priors3}} 
\tablehead{\colhead{Parameter} & 
                  \colhead{CD 35-2722} &
                  \colhead{HIP 23309}     
                  }
\startdata
V$_{radial}$  & U(-250; 250) &  U(-250; 250)  \\
(\kms) & [70.3, 93.3, 99.1, 104.7, 110.1] &  [110.4, 114.5, 118.0, 121.6, 125.1] \\
\hline 
log$_{10}$ A  & U(-18, -8) & U(-18; -8)  \\
(\ergcmsA) & [-12.79, -12.75, -12.70, -12.63, -12.51] & [-12.26, -12.23, -12.20, -12.17, -12.14]  \\
\hline
FWHM$_{L}$  & U(1; 1000) & U(1; 1000)  \\
(\kms) & [162.3, 186.2, 201.5, 215.3, 227.9] &  [122.8, 129.0, 135.6, 142.5, 149.7]  \\
\hline
FWHM$_{G}$  & U(1; 1000) & U(1; 5000)  \\
(\kms) & [236.6, 294.7, 327.9, 359.9, 387.3] &  [421.6, 438.7, 455.3, 472.5, 489.6] \\
\hline
log$_{10}$ N(\HI)  & U(17.5; 19) & U(17.5; 19) \\
(cm$^{-2}$) & [17.52, 17.60, 17.78, 17.98, 18.21] & [17.61, 17.71, 17.80, 17.88, 17.96]  \\
\hline
b$_{HI}$  & 11.5 & 11.5  \\
(\kms) &  &  \\
\hline
V$_{HI}$  & U(-250; 250) & U(-250; 250)  \\
(\kms) & [-66.2, -58.3, 38.2, 59.0, 67.9] & [47.5, 72.0, 74.7, 76.7, 78.8]  \\
\hline
V$_{\rm Si III}$ & U(-250; 250) & U(-250; 250) \\
 & [-7.5, 8.7, 25.1, 42.1, 67.9] & [96.3, 107.5, 118.1, 128.5, 138.6]  \\
 \hline
A$_{\rm Si III}$ & U(-16; -13) & U(-16; -13) \\
 & [-14.28, -14.23, -14.17, -14.11, -14.03] & [-13.86, -13.82, -13.77, -13.72, -13.64]  \\
 \hline
FWHM$_{\rm Si III}$ & U(1; 700) & U(1; 700)\\
 & [338.8, 409.7, 477.8, 548.6, 616.4] & [201.1, 243.0, 276.1, 306.9, 336.5] \\
\hline
\hline
F(\Lya)  & [2.30, 2.37, 2.50, 2.72, 2.99] & [5.03, 5.21, 5.40, 5.60, 5.82]   \\
(erg cm$^{-2}$ s$^{-1}$) & $\times$10$^{-13}$ & $\times$10$^{-13}$ \\
F(Si III)  & [1.26, 1.32, 1.37, 1.43, 1.49] &  [1.85, 1.93, 2.01, 2.10, 2.17] \\
(erg cm$^{-2}$ s$^{-1}$) & $\times$10$^{-14}$ & $\times$10$^{-14}$  \\
\enddata
\tablecomments{U represents a uniform prior within the bounds. Other values are fixed values. On the second line: [2.5\%, 15.9\%, 50\%, 84.1\%, 97.5\%].}
\end{deluxetable}

\begin{deluxetable}{ccc}
\tablecolumns{3}
\tablewidth{0pt}
\tablecaption{ Prior Probabilities, Best Fits, and Confidence Intervals for HIP 112312 (E140M) \label{table:priors_hip112312}} 
\tablehead{\colhead{Parameter} & 
                  \colhead{Native Resolution} &
                  \colhead{Degraded (G140L) Resolution}
                  }
\startdata
V$_{radial}$  & U(-250; 250) & U(-100; 100)    \\
(\kms) & [-3.2, -2.2, -1.1, 0.0, 1.1] &  [16.1, 23.2, 31.6, 39.3, 46.1] \\
\hline 
log$_{10}$ A  & U(-18; -8) & U(-18; -8) \\
(\ergcmsA) & [-11.15, -11.13, -11.10, -11.07, -11.04] &  [-11.89, -11.82, -11.70, -11.44, -10.81] \\
\hline
FWHM$_{L}$  & U(1; 1000) &  U(1; 1000)\\
(\kms) & [37.8, 40.1, 42.6, 45.1, 47.5] & [44.9, 90.9, 122.9, 143.8, 158.1] \\
\hline
FWHM$_{G}$  & U(1; 1000) & U(1; 1000) \\
(\kms) & [216.3, 221.3, 226.4, 231.6, 236.5] & [4.4, 22.5, 71.5, 127.8, 165.3] \\
\hline
log$_{10}$ N(\HI)  & U(17.5; 19) & U(17.5; 19.0) \\
(cm$^{-2}$) & [18.24, 18.26, 18.28, 18.3, 18.33] & [17.52, 17.61, 17.82, 18.08, 18.39] \\
\hline
b$_{HI}$  & $ln$(5; 20) &  $ln$(5; 20) \\
(\kms) & [10.2, 11.4, 12.2, 12.7, 13.2] & [5.6, 8.3, 13.0, 17.3, 19.4] \\
\hline
V$_{HI}$  & U(-250; 250) & U(-100; 100) \\
(\kms) & [-10.3, -9.9, -9.5, -9.0, -8.6] & [-27.7, -15.1, 0.7, 15.0, 23.5]  \\
\hline
\hline
F(\Lya)  & [2.02, 2.08, 2.14, 2.21, 2.29] &  [1.22, 1.31, 1.48, 1.90, 3.17] \\
(erg cm$^{-2}$ s$^{-1}$) & $\times$10$^{-12}$ & $\times$10$^{-12}$ \\
\enddata
\tablecomments{U represents a uniform prior within the bounds. Other values are fixed values. On the second line: [2.5\%, 15.9\%, 50\%, 84.1\%, 97.5\%].}
\end{deluxetable}

\begin{deluxetable}{ccc}
\tablecolumns{3}
\tablewidth{0pt}
\tablecaption{ Prior Probabilities, Best Fits, and Confidence Intervals for HIP 17695 (E140M) \label{table:priors_hip17695}} 
\tablehead{\colhead{Parameter} & 
                  \colhead{Native Resolution} &
                  \colhead{Degraded (G140L) Resolution}
                  }
\startdata
V$_{radial}$  &  U(-50; 50) & U(-100; 100)   \\
(\kms) & [10.3, 11.4, 12.6, 13.8, 14.9] & [37.9, 45.5, 53.0, 58.4, 62.7] \\
\hline 
log$_{10}$ A  &  U(-18; -8) & U(-18; 8)\\
(\ergcmsA) &  [-11.59, -11.57, -11.54, -11.52, -11.49] & [-11.99, -11.91, -11.74, -11.35, -10.87] \\
\hline
FWHM$_{L}$  &  U(1; 1000) & U(1; 1000) \\
(\kms) &  [54.7, 57.7, 60.9, 64.0, 67.2] & [36.8, 64.0, 98.9, 121.5, 137.3] \\
\hline
FWHM$_{G}$  & U(1; 1000) & U(1; 1000)\\
(\kms) &  [143.2, 149.6, 156.2, 162.9, 169.4] & [19.0, 70.8, 110.0, 151.1, 189.5] \\
\hline
log$_{10}$ N(\HI)   & U(17.8; 18.0) & U(17.5; 19.0)\\
(cm$^{-2}$) &  [17.80, 17.80, 17.81, 17.82, 17.84] & [17.51, 17.55, 17.69, 17.93, 18.19] \\
\hline
b$_{HI}$  & $ln$(5; 20) & $ln$(5; 20) \\
(\kms)  & [11.8, 12.0, 12.2, 12.5, 12.6] & [5.3, 6.8, 10.1, 14.0, 18.0] \\
\hline
V$_{HI}$  & U(-50; 50) & U(-100; 100)\\
(\kms) &  [15.1, 15.5, 15.9, 16.3, 16.8] & [-7.5, 16.3, 32.1, 42.1, 51.1]  \\
\hline
\hline
F(\Lya)  & [1.04, 1.07, 1.10, 1.12, 1.15] & [0.86, 0.93, 1.15, 1.88, 3.50] \\
(erg cm$^{-2}$ s$^{-1}$) & $\times$10$^{-12}$ & $\times$10$^{-12}$ \\
\enddata
\tablecomments{U represents a uniform prior within the bounds. On the second line: [2.5\%, 15.9\%, 50\%, 84.1\%, 97.5\%].}
\end{deluxetable}

\bibliography{FUMES_LyA.bib}{}
\bibliographystyle{aasjournal}

\end{document}